\documentclass[12pt,preprint]{aastex}
\slugcomment{Accepted to AJ}

\shorttitle{$\eta$ Carinae 2009.0 Spectroscopic Event}
\shortauthors{Richardson et al.}

\begin{document}

\title{The H$\alpha$ Variations of $\eta$ Carinae\\
 During the 2009.0 Spectroscopic Event}

\author{N. D. Richardson\altaffilmark{1}, 
D. R. Gies\altaffilmark{1},
T. J. Henry\altaffilmark{1},
E. Fern$\acute{\rm a}$ndez-Laj$\acute{\rm u}$s\altaffilmark{2},
and A. T. Okazaki\altaffilmark{3}}
\altaffiltext{1}{Center for High Angular Resolution Astronomy, 
Department of Physics and Astronomy, 
Georgia State University, P. O. Box 4106, Atlanta, GA  30302$-$4106; 
richardson@chara.gsu.edu, gies@chara.gsu.edu, thenry@chara.gsu.edu} 
\altaffiltext{2}{Facultad de Ciencias Astron$\acute{\rm o}$micas y 
Geof$\acute{\rm i}$sicas, Universidad Nacional de La Plata, 
Observatorio Astron\'{o}mico, Paseo del Bosque s/n, 
La Plata, BA, B1900FWA, Argentina; eflajus@fcaglp.unlp.edu.ar}
\altaffiltext{3}{Faculty of Engineering, Hokkai-Gakuen University, 
Toyohira-ku, Sapporo 062-8605, Japan; okazaki@elsa.hokkai-s-u.ac.jp}

\begin{abstract}
We report on H$\alpha$ spectroscopy of the
2009.0 spectroscopic event of $\eta$~Carinae collected via SMARTS
observations using the CTIO 1.5 m telescope and echelle spectrograph. 
Our observations were made almost every night 
over a two month interval around the predicted minimum of $\eta$~Car.
We observed a significant fading of the line emission that reached 
a minimum seven days after the X-ray minimum.  About 17~d prior
to the H$\alpha$ flux minimum, the H$\alpha$ profile exhibited the 
emergence of a broad, P~Cygni type, absorption component (near a 
Doppler shift of $-500$ km~s$^{-1}$) and a narrow absorption 
component (near $-144$ km~s$^{-1}$ and probably associated with 
intervening gas from the Little Homunculus Nebula).  All these 
features were observed during the last event in 2003.5 and are 
probably related to the close periastron passage of the companion. 
We argue that these variations are consistent with qualitative expectations 
about changes in the primary star's stellar wind that result from 
the wind - wind collision with a massive binary companion and 
from atmospheric eclipses of the companion.  
\end{abstract}


\setcounter{footnote}{3}

\section{Introduction}

The star $\eta$ Carinae (HD~93308) is one of the most massive and luminous 
objects in the local region of the Galaxy.  It experienced a large eruption in 
the mid-nineteenth century that led to the formation of the surrounding 
Homunculus Nebula (Davidson \& Humphreys 1997).  The high-excitation, 
forbidden lines in the spectrum disappear and reappear in a 5.54~y cyclic 
manner (Damineli et al.\ 2000).  This same periodicity is observed in the
star's photometric (van Genderen et al.\ 2003; Fern\'{a}ndez-Laj\'{u}s
et al.\ 2003; Whitelock et al.\ 2004) and X-ray light curves (Corcoran
2005).  Damineli et al.\ (1997) and others proposed that 
this period corresponds to the orbital period of a binary companion
in a very eccentric orbit, in which the spectroscopic event (as well as
photometric and X-ray minima) occurs near periastron.  
Detection of the companion has eluded observers so far, but a fit of 
the spectrum from the {\it Chandra X-ray Observatory} with models of the X-rays 
generated in the wind - wind collision indicates that the companion 
has a powerful stellar wind and is probably also a massive star 
(Pittard \& Corcoran 2002).   X-rays from the wind - wind collision 
region will encounter a varying column density of gas along our 
line of sight with the changing binary orientation. 
The X-ray maximum occurs shortly before periastron when 
our sight line cuts through the rarefied wind of the secondary
(bounded by a Coriolis-deflected, bow shock where the winds collide), 
while the X-ray minimum occurs close to periastron when the collision 
region is blocked by obscuring, dense gas of the primary's wind. 
Numerical models of the wind - wind collision by Okazaki et al.\ (2008) 
and Parkin et al.\ (2009) can reproduce many of the features of 
the X-ray light curve, but they also reveal discrepancies from observations 
made during the intense interaction at closest approach.  
Although the details of wind collision need further investigation, 
the basic geometry of the models helps explain the spatial variations of 
the emission spectrum from the resolved, extended wind region surrounding
the central binary (Gull et al.\ 2009).  

The strong stellar wind and surrounding ejecta of $\eta$~Car produce a very strong
H$\alpha$ $\lambda 6563$ line. It is one of strongest emission features in the spectrum and
is formed over a large volume (Hillier \& Allen 1992; Hillier et al.\ 2001).  
The main emission component is formed in the wind of the primary 
(Hillier et al.\ 2001; Davidson et al.\ 2005; Gull et al.\ 2009), but there are other 
components formed in nearby circumstellar gas, especially the Weigelt blobs 
(Weigelt \& Ebersberger 1986; Davidson et al.\ 2005;  Gull et al.\ 2009). 
There is dust in the ejecta of the Homunculus Nebula that scatters the starlight of the 
primary star so that the spatial variations in the emission profile can be used to sample the
wind as seen from different orientations.  Such investigations indicate that the 
star has a strong and fast polar wind and a slow and dense equatorial outflow
(Smith et al.\ 2003; Stahl et al.\ 2005).  During the 2003.5 event 
Davidson et al.\ (2005) observed a decrease in H$\alpha$ emission strength
and the development of a P~Cygni absorption component.  However, we
have not observed enough periastron events to know if these developments
occur every time.

We collected high dispersion H$\alpha$ spectroscopy of $\eta$~Car during 
the recent 2009.0 event through the 
Small and Moderate Aperture Research Telescope System (SMARTS) program with the 
Cerro Tololo Interamerican Observatory (CTIO) 1.5~m telescope.  We present in 
\S2 the spectroscopic observations made before, during, and after the minimum. 
In \S3 we discuss the observed variations and compare them to those observed 
by Davidson et al.\ (2005) during the last (2003.5) event. In \S4, we argue that 
many of the observed changes are consistent with qualitative predictions of the wind -
wind collision model.

\section{Observations}

We obtained high resolution ($R \simeq 40000$; $0.16$\AA ~FWHM) echelle spectra 
of $\eta$~Car with the 1.5~m telescope at CTIO.  The Fiber Echelle 
Spectrograph\footnote{http://www.ctio.noao.edu/$^\sim$atokovin/echelle/index.html}
is connected to the telescope by a fiber that collects flux over a circular 
aperture of diameter $2\farcs7$ on the sky.  The resulting spectra will include
flux contributions from the central binary, extended wind, Weigelt blobs, 
and some scattered light from dust in the Homunculus Nebula 
(Smith et al.\ 2003; Gull et al.\ 2009).  
Our observations have a good signal-to-noise ratio (between 40 and 100 per pixel 
in the continuum, depending on the order) and continuous coverage 
in the spectral region 4800 $-$ 7400\AA .  We collected our first echelle spectrum 
on 2008 Nov 14, and then began obtaining approximately one spectrum per night 
over the period from 2008 Dec 18 to 2009 Feb 19 (UT). 
We typically obtained three 30~s exposures, short enough to avoid saturation at H$\alpha$, 
the brightest feature in this spectral region.  We also made longer exposure (120~s) 
spectra to obtain higher signal-to-noise for transitions weaker than H$\alpha$,
and we will discuss those spectra in a subsequent paper. 
The observations were recorded on a 2048$\times$2048 SITe chip with 24 $\mu$m pixels. 
The chip is read out using two amplifiers, 
and images are recorded with an overscan region in the central 
part of the chip.  All the spectra were bias-subtracted, trimmed, flat fielded, and 
wavelength calibrated using standard 
IRAF\footnote{IRAF is distributed by the National Optical Astronomy Observatory, 
which is operated by the Association of Universities for Research in Astronomy, Inc., 
under cooperative agreement with the National Science Foundation.} techniques for 
echelle spectroscopy.  Wavelength calibration was achieved by means of a ThAr lamp, 
and the typical residuals of our solution were approximately 0.0035 \AA. 
We extracted 46 orders from each spectrum, ranging from 50 \AA\ to 150 \AA\ of 
usable data in each order.  The H$\alpha$ spectra were normalized to a unit 
continuum after large scale sensitivity variations were removed using 
flat field images, and the final spectra were transformed
to a uniform, heliocentric wavelength grid. 

We also collected some lower resolution Cassegrain spectra using the 
CTIO 1.5~m telescope and R-C Spectrograph 
with the standard 47/Ib setup (Howell et al.\ 2006), which records the range from 
5630 \AA\ to 6950 \AA.  These spectra have a resolution of 2.2 \AA\ FHWM ($R \simeq 3000$). 
The spectrograph slit was $1\farcs5$ wide and $300\arcsec$ in height and was 
oriented in an east - west direction.  The spectra were recorded on a Loral 
1200$\times$800 CCD detector and were reduced using standard IRAF techniques. 
Exposures were only 0.2~s long in order to avoid saturation at H$\alpha$. 
We typically made two integrations per visit and performed the wavelength calibration 
using a Ne lamp.  We also observed the flux standard stars Feige~110 or LTT~4364 
to flux calibrate the spectra, but the spectra were subsequently normalized to 
a unit continuum and transformed to a standard, heliocentric wavelength grid. 

In Table 1 we present the dates of our observations.  We used the period 
of 2022.7~d and epoch of minimum of HJD~2452819.8 derived by Damineli et al.\ (2008a) 
to phase our data to the orbital cycle of the system.  Damineli et al.\ define phase 0.0 
to be the time when the narrow emission component of the optical \ion{He}{1} lines disappears. 
They count the 2003.5 event as the eleventh one since the first event noted
during 1947 by Gaviola (1953).  The predicted time of the 2009.0 event of 
HJD~2454842.5 was confirmed in recent observations by Damineli et al.\ (2009). 
Note that the beginning of the X-ray minimum occurred about 4~d after 
this\footnote{http://asd.gsfc.nasa.gov/Michael.Corcoran/eta\_car/etacar\_rxte\_lightcurve/index.html}
and the $V$-band minimum occurred $\approx 16$~d after phase 0.0 (Fern\'{a}ndez-Laj\'{u}s et al.\ 2009). 

\placetable{tab1}  

\section{H$\alpha$ Observations and Variability During the 2009.0 Spectroscopic Event}

Figure 1 displays the line profiles observed with the echelle spectrograph. 
The main feature is a large broad emission structure that is probably 
formed in the wind of the primary star (Davidson et al.\ 2005; Gull et al.\ 2009). 
There are no large scale changes in line position visible, but 
we measured the radial velocity in the sensitive line wings 
using a line bisector method (Shafter et al.\ 1986).  
This method samples the line wings using oppositely signed 
Gaussian functions and determines the mid-point position 
between the wings by cross-correlating these Gaussians with the profile. 
We used Gaussian functions with FWHM = 20 km~s$^{-1}$ at sample 
positions in the wings of $\pm 300$ km~s$^{-1}$, and the resulting 
bisector velocities $V_b$ are given in column 6 of Table~1. 
These measurements (discussed in the next section) show that  
the broad emission shifted slightly from blue to the red near periastron 
(in the same sense as expected for the orbital motion of the primary star; 
Nielsen et al.\ 2007).  There is a narrow emission line in the central part of the profile that 
originates in the nearby Weigelt blobs (Davidson et al.\ 2005; Gull et al.\ 2009). 
We made no attempt to remove this feature because spatially resolved spectra 
(Gull et al.\ 2009) show that the nebular emission extends over the full
emission profile and because this nebular component probably varies near phase 0.0.  
The spectral sequence also shows the development of a narrow absorption line
near a radial velocity of $-144$ km~s$^{-1}$.  This feature appeared in 
earlier spectra (Melnick et al.\ 1982; Damineli et al.\ 1998; Davidson et al.\ 2005) 
and was referred to as the ``anomalous absorption'' by Humphreys et al.\ (2002).  
Several other narrow but weaker absorption 
features are also seen that also appear to be stationary. The blue side of the 
profile displayed a bumpy appearance just prior to the minimum in a manner reminiscent of 
the emission lines of W-R stars (L\'{e}pine et al.\ 1996), and these may result from 
structure and clumping in the outflow of the primary (Hillier et al.\ 2001).  

\begin{figure}
\label{fig1}
\includegraphics[angle=0,width=5in]{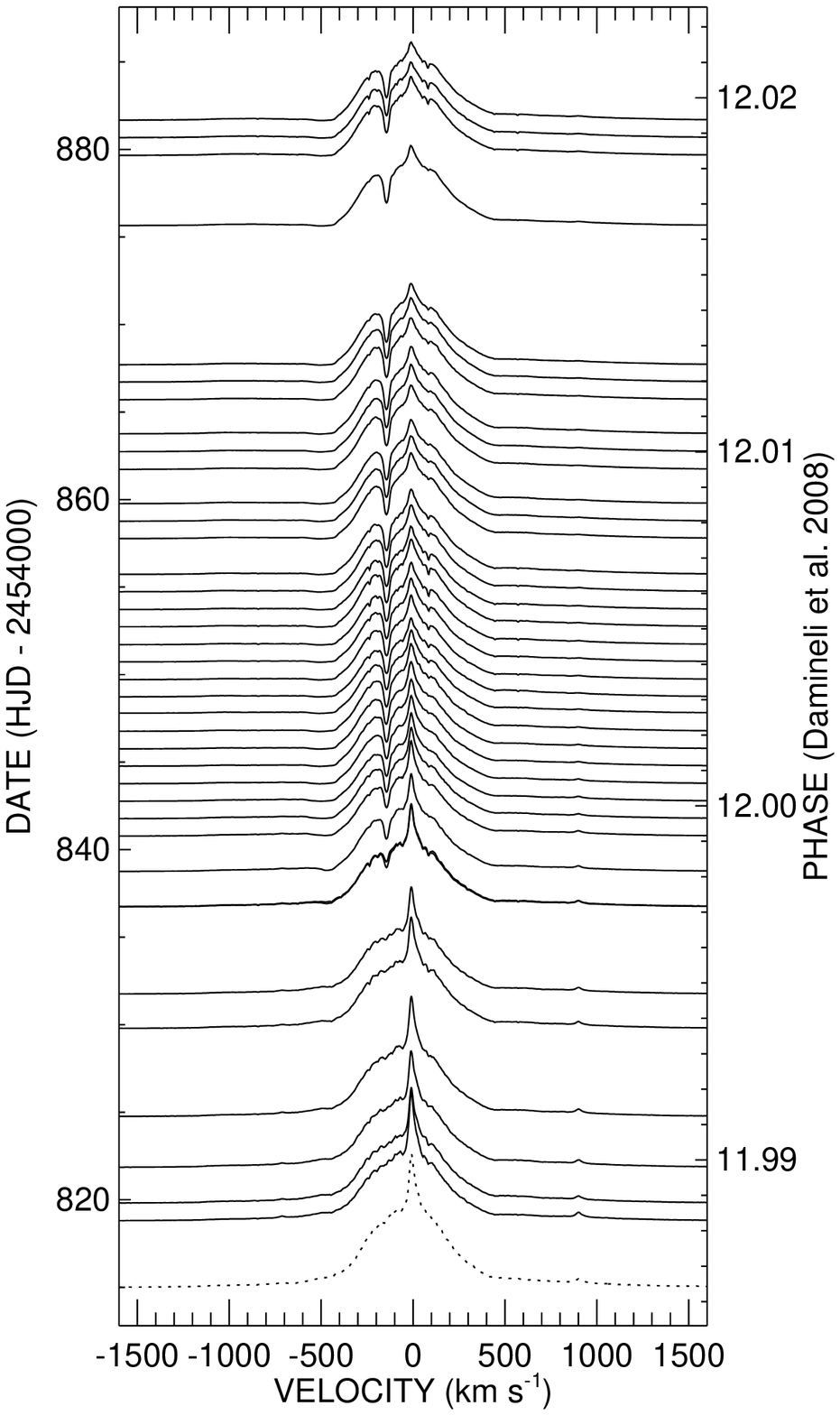}
\caption{A montage of H$\alpha$ profiles of $\eta$ Car
plotted as a function of time and heliocentric
radial velocity.  Each profile is placed so that
the continuum level is set at the heliocentric
Julian date ({\it left}) or orbital phase ({\it right}) 
of observation (Damineli et al.\ 2008a), with the exception 
of the first spectrum ({\it dotted line}) that was made
on HJD 2454784.9 (phase 11.972).  The line intensity
is scaled such that $10\times$ the continuum
strength equals one day of time.}
\end{figure}

\placefigure{fig1} 

Figure 2 shows a plot of the logarithmic intensity of the normalized spectra in a 
gray scale depiction as a function of radial velocity and time.  The benefit of this 
representation is that fainter features in the wings are more readily visible. 
Figure~2 shows how the anomalous absorption feature near $-$144 km s$^{-1}$ and 
a second, P~Cygni type, absorption feature near $-500$ km s$^{-1}$ both 
appear in the H$\alpha$ profile between observations made on 
HJD~2454832 and HJD~2454837 (just prior to phase 0.0 at HJD~2454842.5).  
The P~Cygni absorption shows a slight blueward progression 
of the profile as it becomes stronger that reaches a minimum velocity approximately 
7~d after its first appearance.  At about this time, we see evidence of a 
flat absorption plateau that extends blueward to $-1000$ km~s$^{-1}$ and lasts
for over 10~d.  This velocity range of absorption is similar to that inferred 
for the polar wind of the primary star from observations of scattered light 
spectra in the Homunculus Nebula (Smith et al.\ 2003).   
A similar development of P~Cygni absorption was also 
observed in the 2003.5 event in both H$\alpha$ and higher members of the 
Balmer sequence (Davidson et al.\ 2005; Nielsen et al.\ 2007). 
The [\ion{N}{2}] $\lambda 6583$ emission (near $+900$ km s$^{-1}$ in the H$\alpha$ frame)
is seen throughout the sequence, although its strength is very weak after phase 0.0.

\begin{figure}
\label{fig2}
\includegraphics[angle=0,width=5in]{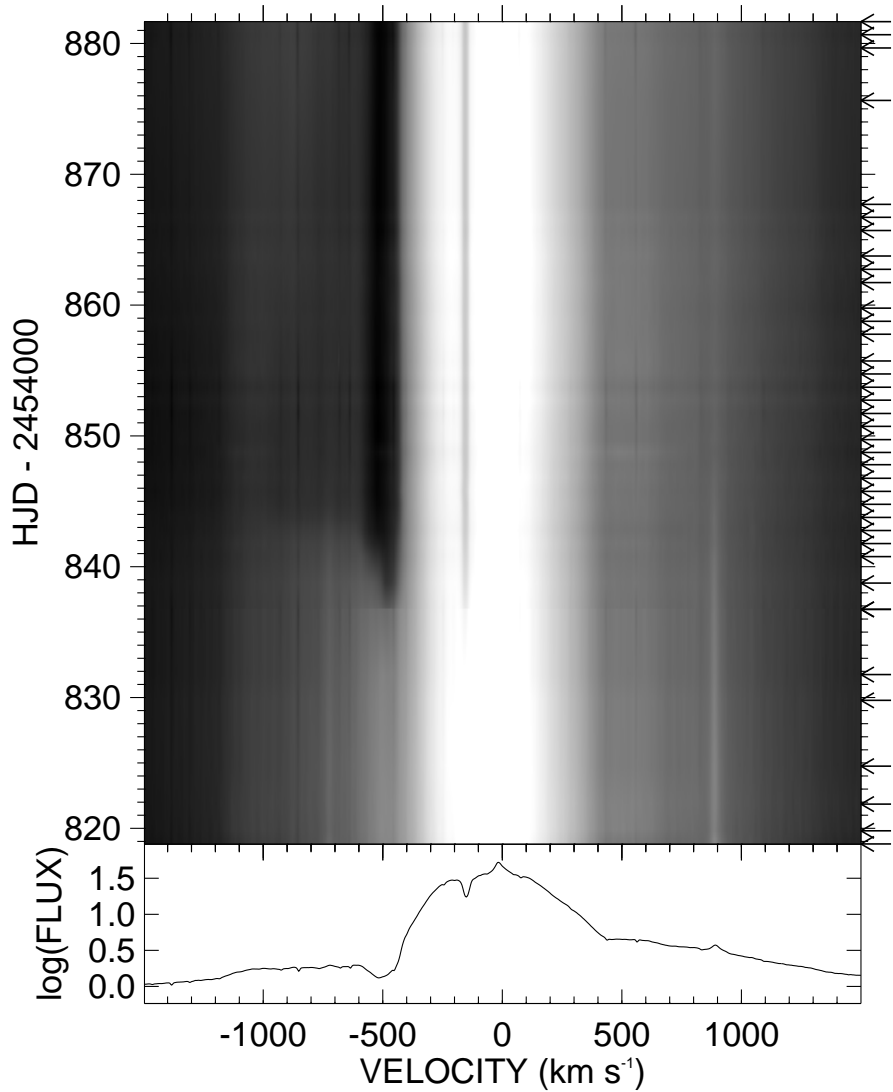}
\caption{A gray scale representation of the H$\alpha$ profiles, 
where the intensity is scaled to the logarithm of the normalized flux. 
The profile in the bottom panel represents the average logarithmic profile. 
Arrows along the right note the actual times of our observations. 
The narrow absorption at $-144$ km~s$^{-1}$ and the P~Cygni absorption at
$-500$ km~s$^{-1}$ both appear to strengthen near HJD~2454837, just prior
to phase 0.0.}
\end{figure}

\placefigure{fig2} 

We compare in Figure~3 the variations observed in 2009.0 with those 
from the prior 2003.5 event.  The H$\alpha$ spectra from the 2003.5 minimum
are {\it Hubble Space Telescope} Space Telescope Imaging Spectrograph (STIS)
observations made available at the HST Treasury Program on Eta Carinae
archive site\footnote{http://etacar.umn.edu/archive/}.  
The STIS spectra were summed across the entire slit to include 
the surrounding nebulosity and to make them more comparable to our ground-based data.
We show the STIS spectra ({\it left panel}) together with our extended set of low 
resolution R-C spectra augmented with several echelle spectra ({\it right panel}),
and the STIS and echelle spectra are smoothed to the lower resolution of the 
R-C spectra for ease of comparison.  The H$\alpha$ profiles are different
between these two events (partly because the space- and ground-based 
observations sample different parts of the inner nebula and possibly due
to the clumpy nature of the primary's wind; Hillier et al.\ 2001), 
but they do appear to show similar trends. 
They both exhibited minimum strength near phase 0.0 and showed the
development of the P~Cygni absorption near $-500$ km~s$^{-1}$ around 
this time (see Fig.~3 in Davidson et al.\ 2005). 
Furthermore, the anomalous absorption near $-144$ km~s$^{-1}$ appears
to strengthen around phase 0.0 in both data sets.  This suggests that 
these features are modulated on the orbital cycle. 

\begin{figure}
\label{fig3}
\includegraphics[angle=90,width=6in]{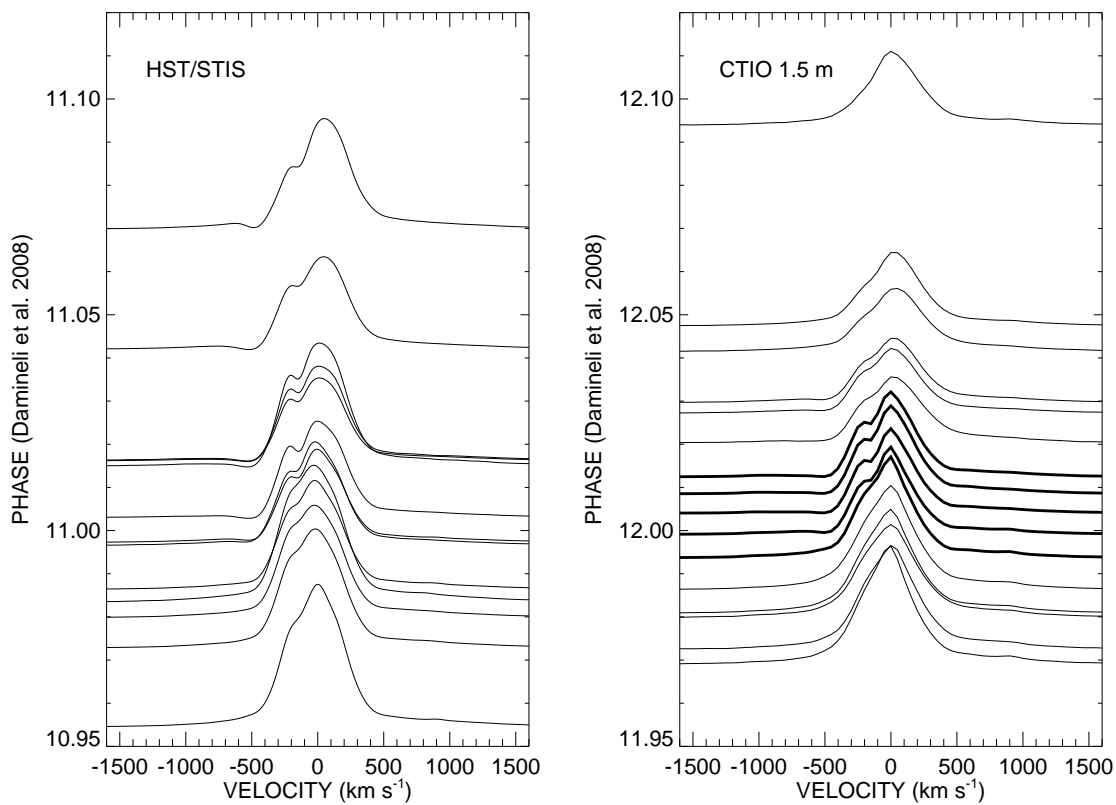}
\caption{A montage of H$\alpha$ profiles of $\eta$ Car as it progressed 
through the last two spectroscopic events. The 2003.5 event spectra ({\it left}) 
are from a summation along the slit of the {\it HST} STIS spectra.  
The 2009.0 event spectra ({\it right}) are from CTIO (all the R-C spectra 
plus five representative spectra from the echelle set ({\it thick lines}) at mid-event).  
The STIS and echelle spectra were smoothed to the resolution of 
the R-C spectra for ease of comparison. The line intensity
is scaled such that $10\times$ the continuum strength equals 0.005 in phase.}
\end{figure}

\placefigure{fig3} 

We measured the H$\alpha$ equivalent width $W_\lambda$ by integrating the line intensity
over a range corresponding to Doppler shifts of $\pm 2500$ km~s$^{-1}$. 
The measurements are listed in column~4 of Table~1 and are plotted as a function
of time in the top panel of Figure~4.  The typical equivalent width error is $\pm 1 \%$.
We transformed these equivalent width measurements 
to a relative line flux using the $\triangle V$-band light curve of \citet{fer10} 
(and subsequent photometry; see lower panel of Fig.~4) and the relation
$$W_{\lambda, corr} = W_\lambda ~10^{-0.4(\triangle V(t) - \triangle V(t_0))}$$
where the fiducial time was set for that of the first echelle spectrum,
$t_0 =$ HJD~2454784.9.   When the equivalent widths are corrected 
for the changing continuum (given in column 5 of Table 1), 
the minimum of the H$\alpha$ flux occurs about 11~d after phase 0.0 (see middle panel of Fig.~4). 
This time lag is similar to that observed in 2003.5 (see Fig.~4 in Davidson et al.\ 2005).  
The H$\alpha$ strength \citep{dav05} and $V$-band flux \citep{fer10} are generally stronger at 
other orbital phases and are subject to irregular and long term fluctuations. 

\begin{figure}
\label{fig4}
\includegraphics[angle=0,width=5in]{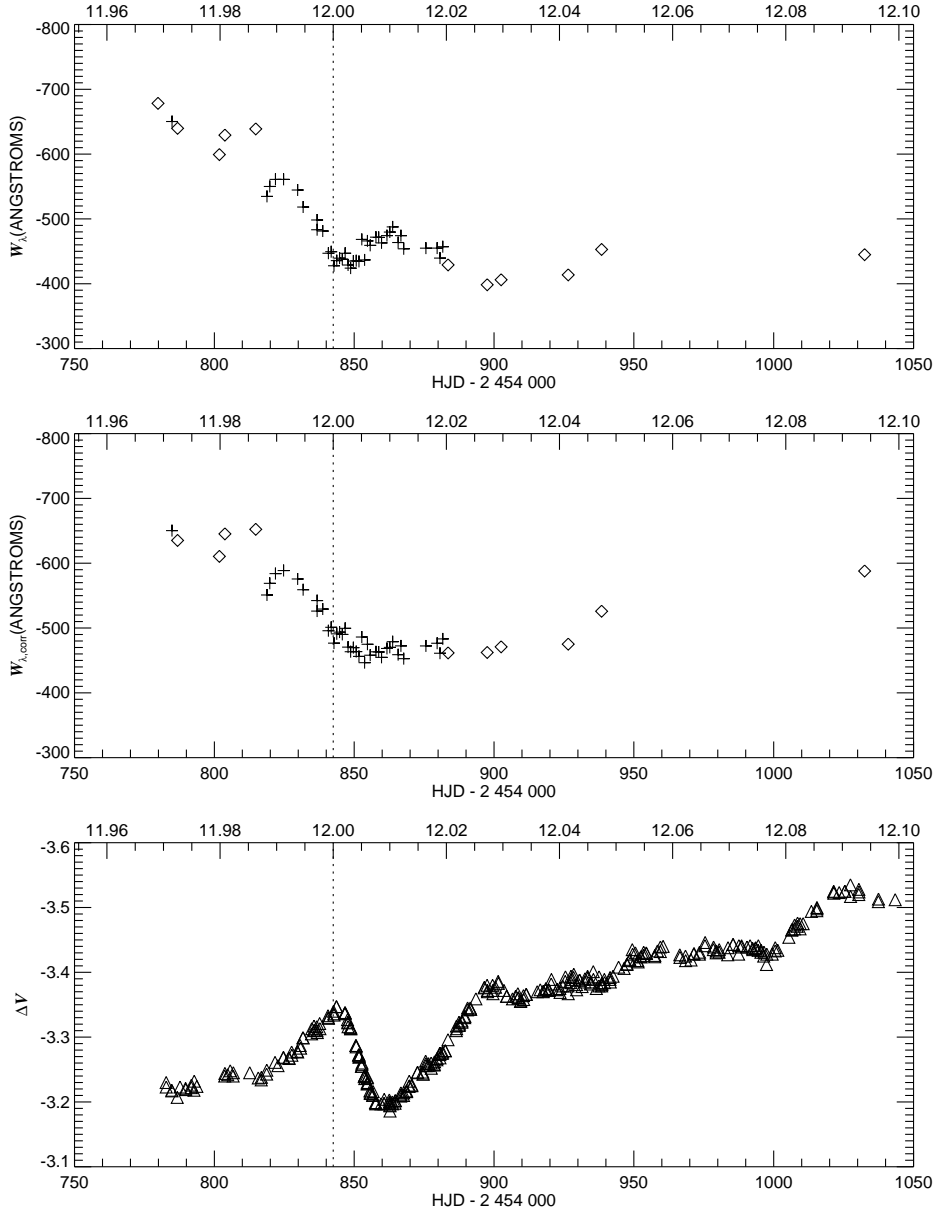}
\caption{The temporal variations of H$\alpha$ strength and $V$-band magnitude. 
In each plot the abscissa represents the date ({\it bottom}) or phase ({\it top}). 
Phase 0.0 from \citet{dam08a} is marked in each plot with a vertical dotted line. 
The top plot shows the equivalent width derived from spectra obtained with 
the R-C spectrograph ($\diamond$) and with the echelle spectrograph ($+$). 
The middle plot shows the equivalent width corrected for the changing continuum
flux that is documented by the differential $V$-band photometry in the lower plot
\citep{fer10}. The $V$-band photometry shows that an eclipse-like 
event began shortly after phase zero and was followed by a general increase
in brightness.}
\end{figure}

\placefigure{fig4} 

\section{Discussion}

Our observations show that some of the H$\alpha$ variations are 
repeatable from event to event, and consequently it is interesting 
to consider what aspects of the binary interaction might be the cause. 
Here we argue that several of the important variations can be
explained in the context of the wind - wind collision model
(Okazaki et al.\ 2008; Parkin et al.\ 2009; Madura et al.\ 2009). 
Okazaki et al.\ (2008) present a hydrodynamical simulation of 
the gas dynamics of the wind - wind collision in $\eta$~Car that 
they use to model the X-ray light curve.  They assume the 
binary consists of a $90 M_\odot$ primary with a slow dense wind
and a $30 M_\odot$ secondary with a fast rarefied wind.  
The orbital eccentricity is set to a large value of $e=0.90$, and
the best fit of the X-ray light curve occurs for a 
longitude of periastron of the primary of $\omega = 243^\circ$ 
and orbital inclination of $i=45^\circ$.  In this configuration, 
the companion spends most of the time near its apastron position
in the foreground along our line of sight (with a separation of roughly 29~AU).
As periastron approaches (and the separation drops to 1.5~AU),
the secondary passes beyond the plane of the 
sky, so that closest approach occurs when the secondary lies 
beyond the primary from our point of view.   
We show in Figure~5 three frames from an animation of one 
simulation\footnote{http://harmas.arc.hokkai-s-u.ac.jp/$^\sim$okazaki/cwb/eta\_car/index.html}. 
This portrays the wind density in the orbital plane for an 
isothermal model.  The left frame shows a time near apastron when the 
secondary ({\it right}) is well separated from the primary ({\it left})
and a well-defined bow shock occurs at the wind - wind collision boundary. 
The dark line in the panel shows our direction of viewing (which 
lies $45^\circ$ below the plane of the diagram).  The middle panel 
shows the configuration 21~d before periastron when the stars are 
much closer and the bow shock is situated very close to the primary. 
Finally the right panel shows the situation after periastron 
(a brief 42~d later) when the companion has moved $188^\circ$ 
counter-clockwise around the primary.  

\begin{figure}
\label{fig5}
\includegraphics[angle=0,width=6in]{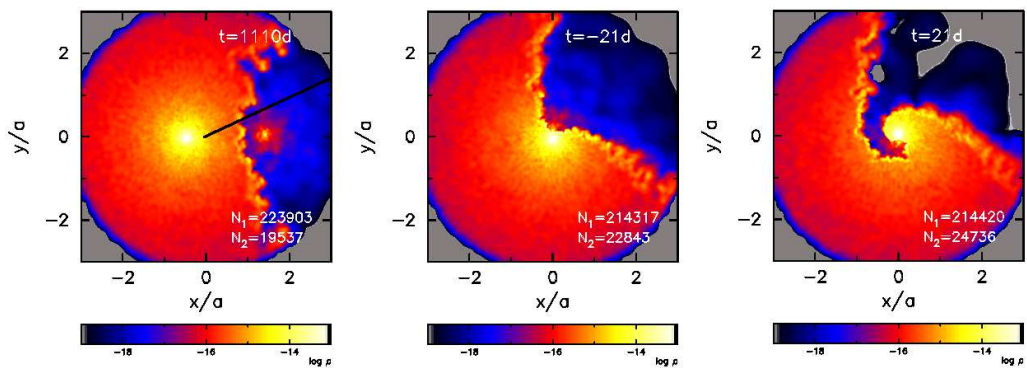}
\caption{Isothermal models of the colliding winds in the binary system 
from the simulations by Okazaki et al.\ (2008).  Each panel shows a density
map in the orbital plane (in spatial units of the semi-major axis). 
The left panel shows the primary (surrounded by its wind; left side) 
and the secondary (dot on right side) at maximal orbital separation. 
We expect the H$\alpha$ flux to form mainly in the densest regions of the wind. 
Our assumed line of sight is indicated by the black line in the left panel
(inclined by $45^{\circ}$ from below the plane of the figure).  The time in days 
relative to periastron is given in the upper right of each panel, and 
the diagrams show how the colliding winds change the density distribution
from the usual situation near apastron ({\it left}) to that at times just before 
({\it middle}) and after ({\it right}) periastron.}
\end{figure}

\placefigure{fig5} 

Madura et al.\ (2009) used similar hydrodynamical simulations to explore 
how the interaction affects the continuum light of the system when the 
stars are closest at periastron.  They find that the fast wind of the 
secondary excavates a region of the dense wind of the primary.  
This ``bore hole'' effect first exposes to view the hotter and deeper 
wind layers and results in an episode of brightening.  However, as the 
secondary progresses past periastron towards superior conjunction, 
the cavity region turns away from us and the system fades.  The drop 
in flux continues past superior conjunction as the cavity reduces the 
effective volume of wind gas, and the system slowly returns to its 
original flux as the stars separate and wind material refills the cavity.

Let us now reconsider the H$\alpha$ orbital variations in the context 
of these wind - wind collision models.  The H$\alpha$ emission forms in dense regions 
of wind (because the emission process is dependent on density squared) that are 
mainly found relatively close to the primary star (Hillier et al.\ 2001).  
As the companion approaches periastron, the bow shock from the colliding winds 
gets closer to the primary, and the developing cavity reduces the effective 
volume of the wind emission from the primary.  The resulting H$\alpha$ emitting 
volume will reach a minimum at periastron, but will then increase again as the 
binary separation increases and the wind is restored to the cavity region. 
Hillier et al.\ (2001) present a model for the wind of the primary star alone in which
the continuum forming radius is approximately 3 AU (where the Rosseland optical 
depth is one) and the H$\alpha$ formation peaks near a radius of 18 AU (see their 
Fig.~15).  Based upon the momentum balance of the colliding winds, the 
wind - wind collision zone will occur at a distance from the primary of $0.67 a(t)$, 
where $a(t)$ is the binary separation at time $t$, and this reaches a minimum of 1~AU in 
the binary model (Okazaki et al.\ 2008).   Thus, the colliding winds cavity should 
significantly reduce the volume of both the continuum and H$\alpha$ emitting regions
close to periastron.  The fact that both the length and relative amplitude of the 
periastron variation is larger for H$\alpha$ than for the continuum (Fig.~4) is consistent
with the larger formation radius of the H$\alpha$ emission.   If we assume that 
periastron occurred when the H$\alpha$ flux was a minimum at HJD 2,454,854 $\pm 2$ 
(the same periastron time within errors assumed by Madura et al.\ 2009 based on a fit
of the continuum light curve), then the H$\alpha$ flux began to decrease about 
39 d before periastron (Fig.~4) when the model distance from the primary to the wind-wind 
shock was 3 AU.  This is closer than the 18 AU H$\alpha$ formation radius given by 
Hillier et al.\ (2001), but we suspect that a marked H$\alpha$ reduction will 
only be observed once a significant fraction of the wind volume is carved out by the 
colliding winds (i.e., after the bow shock apex reaches to relatively deep within the wind). 
After periastron, as the colliding winds zone moves outward, the primary's wind will
re-occupy the volume opened by the fleeing cavity.  Hillier et al.\ (2001) note that the 
wind flow time to the H$\alpha$ formation region is about 100 d, which is comparable 
to the H$\alpha$ flux recovery time that we observed (Fig.~4).   

It is probable that the periastron interaction is also the cause of the H$\alpha$ 
radial velocity variations observed at that time.   We show in Figure~6 the measured 
H$\alpha$ bisector radial velocity $V_b$ (Table~1) together with the expected Keplerian 
orbital velocity curve for the primary.  Although there is a superficial resemblance, 
the observed curve has a much lower semiamplitude than expected.   A similar change
from blue-shift to red-shift around periastron has been reported for other lines 
but usually with a much larger amplitude than the orbital value (Nielsen et al.\ 2007;
Damineli et al.\ 2008b).  We doubt that the velocity shift observed in H$\alpha$ is 
related to Keplerian motion, because the formation radius of the H$\alpha$ flux is so 
large that any orbital motion vanishes due to conservation of angular momentum 
(i.e., all the motion becomes radial).  However, we suspect that the close 
interaction at periastron may have a tidal influence on the wind of the primary 
causing the wind to become locally denser and slower along the instantaneous 
axis between the stars (a ``focused wind''; Friend \& Castor 1982).  Since 
the H$\alpha$ flux varies as density squared, this wind density enhancement would 
result in a slight excess in blue-shifted flux while the companion approaches from
the near side of the plane of the sky and then a red-shifted excess around periastron 
as the companion moves beyond the plane of the sky.  A full radiative transfer calculation 
will be required to test whether or not such a tidal enhancement can explain the timing 
and amplitude of the observed radial velocity variation. 

\begin{figure}
\label{fig6}
\includegraphics[angle=90,height=12cm]{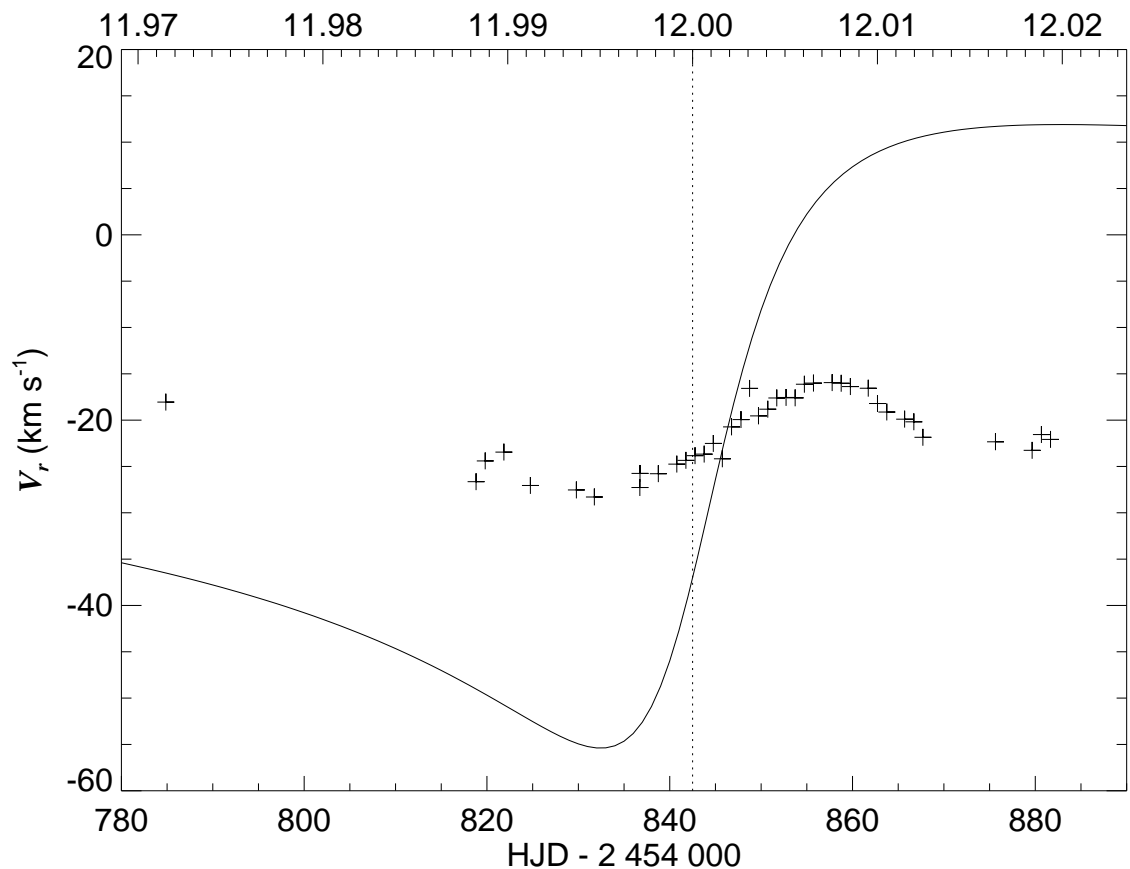}
\caption{The H$\alpha$ bisector velocities $V_b$ ({\it plus signs}) 
plotted against heliocentric Julian date ({\it bottom}) and 
orbital phase ({\it top}).  Phase 0.0 is marked 
with a vertical dotted line.  The symbol sizes are comparable to 
the velocity measurement errors ($\approx 1.0$ km~s$^{-1}$). 
The solid line shows the predicted Keplerian orbital velocity 
curve for the primary star according to the elements adopted by
Okazaki et al.\ (2008), epoch of periastron from Damineli et al.\ 
(2008a; assumed identical to their phase 0.0), and the systemic 
velocity from Smith (2004).}
\end{figure}

\placefigure{fig6} 

The development of a P~Cygni type absorption feature near a radial velocity 
of $-500$ km~s$^{-1}$ around the time of periastron (Fig.~2) was also observed 
by Davidson et al.\ (2005) during the last two events, so it is probably 
related to the binary interaction.  We suggest that the emergence of this absorption 
component results from a change in illumination by the hot companion.  In the binary 
model, the hot companion resides in the foreground during most of the orbit, 
and the radiation from the companion probably tends to overionize the outer 
layers of the hemisphere facing us.  However, as the binary approaches periastron, 
the companion will appear to set below the optically thick horizon 
as seen from a position in the outer layers of the wind facing our direction.  
The subsequent gas cooling will result in greater absorption by this wind material seen 
projected against the primary and will cause an increase in the blue absorption
of the P~Cygni line.  Ironically, it is then during this brief time around 
periastron that we see the undisturbed, ``normal'' wind of the primary 
(at least in the sector directed towards us) with blue-shifted absorption 
lines that are more similar to the model predictions (Hillier et al.\ 2001)
than observed at the other orbital phases when the hemisphere facing us
is illuminated by the secondary star. Through scattered light spectroscopic studies 
of the Homunculus Nebula, Smith et al.\ (2003) found that the polar wind of the primary 
star exhibits a broad absorption plateau on the blue wing of the P Cygni profile.
We see this same behavior in our data (Fig.~2), which suggests that we observe a less
obstructed view of parts of the primary star's wind during the spectroscopic event.

We suspect that this same change in secondary illumination is the cause of 
the sudden appearance of the anomalous, narrow absorption feature 
just prior to periastron (Fig.~2).  This same increase in absorption strength was 
first observed during the 1981.5 event by Melnick et al.\ (1982), then again 
during the 1992.5 event by Damineli et al.\ (1998), and finally during the 
2003.5 event (Fig.~3) by Davidson et al.\ (2005), so the phenomenon is 
clearly linked to the binary clock.    
This narrow absorption feature is probably formed farther out from the 
central binary because the feature appears in the spatially resolved, extended emission 
(see Fig.~15 of Gull et al.\ 2009) where the gas may be cooler (consistent with the 
fact that the anomalous absorption is weak or absent in the profiles of the upper 
Balmer lines; Weis et al.\ 2005; Nielsen et al.\ 2007). 
The feature has the same width and radial velocity as nebular absorption lines that 
are observed in the far-ultraviolet spectrum and that form in the Little Homunculus,
an ionized gas cloud deep within the larger Homunculus Nebula 
(Ishibashi et al.\ 2003; Smith 2005; Gull et al.\ 2006). 
Since our view to central binary passes through the nearby wall of the 
Little Homunculus (see Fig.~2 of Nielsen et al.\ 2009), we suggest that the
H$\alpha$ anomalous absorption forms there.  In the weeks prior to periastron, 
this part of the Little Homunculus is probably exposed to the ionizing 
radiation that escapes through the rarefied gas of the secondary star's wind
and surrounding bow shock that is oriented in our direction at that time. 
We suspect that this ionizing flux leads to a net reduction in the H$\alpha$ absorption 
then because the hydrogen gas in the part of the Little Homunculus along our
line of sight becomes over-ionized.  However, as the secondary progresses towards 
periastron and the bow shock surface passes beyond our view to the 
primary, this source of ionizing flux would decrease suddenly
(effectively blocked by the wind of the primary), and 
the net H absorption would then become much more effective. 
Such a decrease in the far ultraviolet flux was observed during  
the 2003.5 event by Iping et al.\ (2005), who argued that the 
flux decline resulted from an atmospheric eclipse of a hot companion.
The fact that the H$\alpha$ absorption features at velocities of 
$-500$ km~s$^{-1}$  and $-144$ km~s$^{-1}$ both appear at about the 
same time (Fig.~2) strongly supports the idea that both result 
from cooling associated with the temporary disappearance of the 
ionizing flux of the companion star.  Although the wall of the Little 
Homunculus where the absorption forms lies about 16 light days closer
to us than the primary star (Nielsen et al.\ 2009), the wall gas would 
``see'' the drop in ionizing flux at about the same time that light 
arrives from the outer layers of the primary, 
so the increased absorption from the Little Homunculus will appear to us to 
occur at about the same time as that from the primary's wind as long 
as the cooling timescale in the Little Homunculus is short (less than a day). 
This change in illumination probably also explains the apparent
temperature drop in the excitation/ionization properties of 
iron-group element absorption lines from the Little 
Homunculus around the time of periastron (Gull et al.\ 2006). 
 
These arguments suggest that the primary features of the H$\alpha$ emission  
form in the extended wind of the primary that becomes truncated 
and asymmetrical around periastron.  The reality is certainly more 
complicated, and, in particular, the wind collision models predict that high gas 
densities should occur in the vicinity of the bow shock near periastron. 
Given the density squared dependence of the H$\alpha$ emission, it is 
somewhat surprising that we find no clear evidence of H$\alpha$ emission 
from the bow shock region itself (where we might expect 
to find a component with a large radial velocity excursion as the 
bow shock quickly rotates around the primary at periastron).  We suspect 
that the absence of a bow shock emission component may be due to 
the higher gas temperatures there (which would cause a decrease in 
emissivity; Ferland 1980) and/or the presence of intervening, dense gas of 
the primary star's stellar wind that is too optically thick for the 
embedded bow shock flux to escape in our direction. 

Our results support qualitatively the colliding winds model and, in 
particular, the predicted changes in the primary's emitting wind structure that 
occur around periastron.  A detailed comparison of the model predictions will 
require a 3D radiative transfer calculation of the H$\alpha$ line intensities,
a very computationally challenging task.  Observations around periastron of 
other emission lines that form closer to the primary will also provide a 
good test of the validity of the basic picture of the system and its colliding winds. 
The 2009.0 event of $\eta$~Car has offered us the means to begin this kind 
of promising investigation.  In a subsequent paper, we plan to examine other
emission lines in the echelle data and to study their temporal variations 
around periastron.  The timing and duration of line flux reductions for 
transitions formed at different radii will help us test the model of the 
migration of the cavity to inner regions of the primary star's wind and will place 
observational constraints on the wind geometry and structure of $\eta$ Carinae.

\acknowledgments 
We thank Kris Davidson, Ted Gull, Roberta Humphreys, Tom Madura, Krister Nielsen, 
Julian Pittard, and an anonymous referee for their constructive comments.  
These spectra were collected with the CTIO 1.5~m telescope, 
which is operated by the SMARTS Consortium.
We are extremely grateful to Fred Walter (Stony Brook University) for 
his careful scheduling of this program and to the CTIO SMARTS staff for 
queue observing support.   
This research has made use of the data archive for the 
HST Treasury Program on Eta Carinae (GO 9973) which is available 
on-line at http://etacar.umn.edu.  The archive is supported by the 
University of Minnesota and the Space Telescope Science Institute under contract with NASA. 
This work was supported by the National Science Foundation under grant AST-0606861. 
Institutional support has been provided from the GSU College of Arts and Sciences and 
from the Research Program Enhancement fund of the Board of Regents of the University 
System of Georgia, administered through the GSU Office of the Vice President for Research.

\clearpage

\clearpage

\clearpage

\clearpage 

\clearpage

\clearpage

\clearpage

\begin{deluxetable}{cccccc}
\label{tab1}
\tablecaption{H$\alpha$ Measurements}
\tablewidth{0pt}
\tablehead{
	\colhead{Date}	       &
	\colhead{Orbital}	&	
	\colhead{}	       &	
	\colhead{$W_{\lambda}$ (H$\alpha$)}	&
	\colhead{$W_{\lambda, corr}$ (H$\alpha$)}  &		
	\colhead{$V_b$ (H$\alpha$)}  \\	
	\colhead{(HJD - 2450000)}	       &
	\colhead{Phase}	&	
	\colhead{Spectrograph}	       &	
	\colhead{(\AA )}	&
	\colhead{(\AA )}  &		
	\colhead{(km s$^{-1}$)} }		
\startdata
4779.8615	&	11.969	&	R-C	&	$-$678	&    \nodata	& \nodata \\
4784.8577	&	11.972	&	echelle	&	$-$650	&	$-$650	& $-18.1$ \\
4786.8458	&	11.972	&	R-C	&	$-$640	&	$-$635	& \nodata \\
4801.7878	&	11.980	&	R-C	&	$-$599	&	$-$611	& \nodata \\
4803.7731	&	11.981	&	R-C	&	$-$629	&	$-$645	& \nodata \\
4814.8012	&	11.986	&	R-C	&	$-$639	&	$-$652	& \nodata \\
4818.8060	&	11.988	&	echelle	&	$-$535	&	$-$551	& $-26.6$ \\
4819.8108	&	11.989	&	echelle	&	$-$550	&	$-$569	& $-24.4$ \\
4821.8520	&	11.990	&	echelle	&	$-$561	&	$-$584	& $-23.4$ \\
4824.7536	&	11.991	&	echelle	&	$-$561	&	$-$589	& $-27.1$ \\
4829.7838	&	11.994	&	echelle	&	$-$545	&	$-$576	& $-27.5$ \\
4831.7530	&	11.995	&	echelle	&	$-$518	&	$-$559	& $-28.3$ \\
4836.7315	&	11.997	&	echelle	&	$-$483	&	$-$526	& $-27.3$ \\
4836.7444	&	11.997	&	echelle	&	$-$498	&	$-$542	& $-25.8$ \\
4838.7496	&	11.998	&	echelle	&	$-$481	&	$-$529	& $-25.8$ \\
4840.7721	&	11.999	&	echelle	&	$-$447	&	$-$496	& $-24.7$ \\
4841.7684	&	12.000	&	echelle	&	$-$449	&	$-$501	& $-24.3$ \\
4842.7673	&	12.000	&	echelle	&	$-$428	&	$-$477	& $-23.8$ \\
4843.7668	&	12.001	&	echelle	&	$-$436	&	$-$491	& $-23.7$ \\
4844.7694	&	12.001	&	echelle	&	$-$439	&	$-$494	& $-22.5$ \\
4845.7581	&	12.002	&	echelle	&	$-$438	&	$-$490	& $-24.2$ \\
4846.7611	&	12.002	&	echelle	&	$-$447	&	$-$500	& $-20.7$ \\
4847.7971	&	12.003	&	echelle	&	$-$429	&	$-$471	& $-19.9$ \\
4848.7348	&	12.003	&	echelle	&	$-$424	&	$-$463	& $-16.6$ \\
4849.7148	&	12.004	&	echelle	&	$-$435	&	$-$470	& $-19.5$ \\
4850.7272	&	12.004	&	echelle	&	$-$435	&	$-$464	& $-18.8$ \\
4851.7129	&	12.005	&	echelle	&	$-$434	&	$-$456	& $-17.6$ \\
4852.7313	&	12.005	&	echelle	&	$-$468	&	$-$486	& $-17.6$ \\
4853.7222	&	12.006	&	echelle	&	$-$437	&	$-$447	& $-17.6$ \\
4854.7335	&	12.006	&	echelle	&	$-$466	&	$-$475	& $-16.1$ \\
4855.7283	&	12.007	&	echelle	&	$-$459	&	$-$458	& $-16.0$ \\
4857.7714	&	12.008	&	echelle	&	$-$472	&	$-$463	& $-16.0$ \\
4858.7605	&	12.008	&	echelle	&	$-$471	&	$-$463	& $-16.0$ \\
4859.7681	&	12.009	&	echelle	&	$-$463	&	$-$455	& $-16.4$ \\
4861.7205	&	12.010	&	echelle	&	$-$474	&	$-$468	& $-16.6$ \\
4862.7332	&	12.010	&	echelle	&	$-$480	&	$-$470	& $-18.2$ \\
4863.7529	&	12.011	&	echelle	&	$-$488	&	$-$479	& $-19.1$ \\
4865.7026	&	12.011	&	echelle	&	$-$464	&	$-$459	& $-19.9$ \\
4866.7104	&	12.012	&	echelle	&	$-$474	&	$-$472	& $-20.2$ \\
4867.7011	&	12.012	&	echelle	&	$-$454	&	$-$453	& $-21.9$ \\
4875.6456	&	12.016	&	echelle	&	$-$455	&	$-$472	& $-22.3$ \\
4879.6545	&	12.018	&	echelle	&	$-$455	&	$-$477	& $-23.3$ \\
4880.6719	&	12.019	&	echelle	&	$-$440	&	$-$461	& $-21.6$ \\
4881.6680	&	12.019	&	echelle	&	$-$457	&	$-$483	& $-22.1$ \\
4883.5776	&	12.020	&	R-C	&	$-$429	&	$-$462	& \nodata \\
4897.5556	&	12.027	&	R-C	&	$-$398	&	$-$462	& \nodata \\
4902.5327	&	12.030	&	R-C	&	$-$406	&	$-$471	& \nodata \\
4926.5314	&	12.042	&	R-C	&	$-$414	&	$-$475	& \nodata \\
4938.4690	&	12.047	&	R-C	&	$-$453	&	$-$526	& \nodata \\
5032.5139	&	12.094	&	R-C	&	$-$445	&	$-$588	& \nodata \\
\enddata
\end{deluxetable}

\end{document}